\documentclass[twocolumn,showkeys,aps,prd]{revtex4}

\usepackage{dcolumn}
\usepackage{graphicx}
\usepackage{amsmath}
\usepackage{amssymb}
\usepackage{xspace}
\usepackage[linktocpage={true}]{hyperref}

\usepackage{color}

\renewcommand{\eqref}[1]{Eq.\,\ref{#1}}

\begin{document}


\title{Constraints on Yield Parameters in Extended Maximum Likelihood Fits}
\author{
Till Moritz Karbach$^{a}$,
Maximilian Schlupp$^{b}$
}
\affiliation{
$^{a}$TU Dortmund, Germany, {\tt moritz.karbach@cern.ch} \\
$^{b}$TU Dortmund, Germany, {\tt maximilian.schlupp@cern.ch}
}
\date{\today}



\begin{abstract}

The method of extended maximum likelihood is a well known concept
of parameter estimation. 
One can implement external knowledge on the unknown parameters 
by multiplying the likelihood by constraint terms. 
In this note, we emphasize that this is also true for yield parameters
in an extended maximum likelihood fit, which is widely used in the
particle physics community.
We recommend a way to generate
pseudo-experiments in presence of constraint terms on yield parameters,
and point to pitfalls inside the {\sc RooFit} framework.

\end{abstract}


\maketitle


\section{Introduction}

The concept of extended maximum likelihood (EML) is widely used for parameter
estimation in particle physics. It is described in~\cite{Barlow},
and we shall summarize its main features here. In EML, the total number of 
events is regarded as a free parameter. Its best value is determined by 
maximizing the likelihood function.
The number of observed events follows a probability density function (PDF),
typically a Poisson PDF. In some situations, the observed number of events
is not the most efficient estimator for the expected number of events. 
These situations occur
when there is at least one free parameter (or a combination of parameters) that simultaneously 
changes both shape and normalization of the PDF.
Then, an EML fit is superior to a regular maximum likelihood (ML) fit. These genuine EML situations
are labeled ``type A'', following the notation of~\cite{Barlow}. 

The textbook example of a type A situation is that of an unknown
signal over a known background of $N_b$ events. Suppose both
signal and background are described by unit Gaussian PDFs \mbox{$G(x;\mu,\sigma=1)$}, then
one possible (non-normalized) total PDF is
\begin{equation}
	\label{eq:typeA}
	g(x) = N_s G(x;\mu_1=0) + N_b G(x;\mu_2=0.5)\ ,
\end{equation}
with $N_s$ being the only free parameter.

Besides the genuine EML situation, there are also ``type B'' EML
situations (or ``bogus'', following again~\cite{Barlow}), where both EML and ML give equivalent
results. This is the case when in \eqref{eq:typeA} also $N_b$ is
a free parameter. Then we can rewrite
\begin{equation}
	\label{eq:typeB}
	g(x) = N (f G(\mu_1=0) + (1-f) G(\mu_2=0.5))\ ,
\end{equation}
with the total number of events $N=N_s+N_b$ and the signal fraction $f=N_s/N$. Now $f$ controls
only the shape of the PDF, while $N$ controls only the
normalization. It might still be beneficial to formulate a problem
using EML terms as in \eqref{eq:typeA}, even if it truly is a type B
problem. This is because the ML notation from \eqref{eq:typeB}
quickly leads to less intuitive fraction parameters if more than
one background component is present, while the yields of \eqref{eq:typeA}
are interpreted easily.

The extended likelihood is formed by multiplying the classical
likelihood by a Poisson term,
\begin{equation}
	\label{eq:el}
	\mathcal{L}(N,\vec{\lambda}) = 
	\frac{e^{-N} N^{N_{\rm obs}}}{N_{\rm obs}!}
	\times \prod_{i=1}^{N_{\rm obs}} \mathcal{P}(\vec{x}_i;\vec{\lambda})\ ,
\end{equation}
where $N$ and $N_{\rm obs}$ are the number of expected and
observed events, respectively, $\mathcal{P}$ is the total PDF, 
$\vec{x}$ is the vector of observables, $\vec{\lambda}$
is the vector of parameters to be estimated. The constant factorial 
term ($N_{\rm obs}!$) is usually omitted as it does not change the shape
of $-\ln\mathcal{L}$ at its minimum.

In the following we discuss how to include external constraints
into the (extended) likelihood and review the effects of such terms. Then we
describe a way to generate pseudo (``toy'') experiments, and demonstrate, that
it will lead to unbiased results, if the correct pull statistic is
chosen. We will show that this is still the case when constraints on 
yield parameters are present. At last, we will point out several 
pitfalls that are present in the toy experiment tools of a current 
version of the {\sc RooFit} framework.


\section{Constraints}
\label{sec:constraints}

If there is knowledge available on the true value of a fit parameter,
we can incorporate this knowledge into the fit procedure.
For example, a previous experiment might have already measured the
parameter at hand, and we have access to their published result,
say $\lambda_e \pm \sigma_{\lambda_e}$.
It is well known how to incorporate such constraints into maximum 
likelihood fits. The full likelihood function is 
multiplied by the constraint PDF $\mathcal{C}(\lambda)$ (where $\lambda$
be a component of $\vec{\lambda}$)
\begin{equation}
	\mathcal{L}_c = \mathcal{C}(\lambda) \times \mathcal{L}\ .
\end{equation}
This holds also in the EML case, and also for constraints on yield
parameters---even though the likelihood is not Poissonian
anymore in the total yield, but contains the product of a Poisson term
in the total yield and a non-Poissonian constraint term in a component yield.
Often a Gaussian distribution is assumed for $\mathcal{C}(\lambda)$,
\begin{equation}
	\label{eq:constlikelihood}
	\mathcal{C}(\lambda) = \frac{1}{\sqrt{2\pi}\sigma_{\lambda_e}}
	\exp\left(-\frac{(\lambda_e-\lambda)^2}{2\sigma_{\lambda_e}}\right)\ .
\end{equation}
%
If more than one parameter is constrained, there can in principle
be an external correlation between them.
This external correlation is different from the internal one. It can
easily be accounted for by, for example, replacing the single Gaussian
of \eqref{eq:constlikelihood} by a multivariate one,
\begin{equation}
	\label{eq:constmultivar}
	\mathcal{C}(\lambda) = 
	\frac{\exp \left(-(\vec{\lambda}_e-\vec{\lambda}) V_e^{-1} (\vec{\lambda}_e-\vec{\lambda})^T\right)}
	{(2\pi)^{l/2} \sqrt{|V_e|}}\ ,
\end{equation}
where $V_e$ is the known $l\times l$ external covariance matrix.

We now recall two effects of including a constraint term
for parameter $\lambda$: they include external knowledge, and
are a means of error propagation.

Compared to a situation with a floating
parameter and no constraint, including the constraint term will reduce
the reported error on this parameter. Suppose that when $\lambda$ is left 
floating without constraint, the result be $\lambda_u \pm \sigma_{\lambda_u}$,
and with the constraint term included it be $\lambda_c \pm \sigma_{\lambda_c}$. 
If both the unconstrained likelihood and the constraint term are 
uncorrelated and Gaussian in $\lambda$, 
the likelihood fit is equivalent to the weighted average of $\lambda_u$
and $\lambda_e$. Thus the error will be given by
\begin{equation}
	\label{eq:weightedaverr}
	\frac{1}{\sigma_{\lambda_c}^2} = \frac{1}{\sigma_{\lambda_u}^2} + \frac{1}{\sigma_{\lambda_e}^2}
\end{equation}
so that $\sigma_{\lambda_c} < \sigma_{\lambda_u}$.

%

Constraint terms are also a means of error propagation.
If the likelihood depends not only on the fit parameters, but also on
parameters that are fixed, one may want to propagate the errors of the
fixed parameters into the fit result. This can be done by including
constraint terms in the fixed parameters, and letting the previously fixed 
parameters float, too.
If there are non-zero correlations between the previously fixed and the 
floating parameters, the errors on the latter will increase, reflecting
the propagated uncertainty on the previously fixed parameters. The reported errors
on the previously fixed parameters will in general be smaller than given by the
constraint. This is because the dataset can also hold information on them.

In addition to the above effects, 
constraint terms can also be incorporated to help the fit
converge. When doing this, the errors are modified, for example as indicated
by \eqref{eq:weightedaverr}. This might spoil the interpretation of the
reported fit errors as being ``statistical'', if $\sigma_{\lambda_e}$ is
not statistical and also of same order as $\sigma_{\lambda_u}$.

The effects of constraints described above are not limited to
shape parameters. They also apply to normalization parameters
such as the fraction parameters of~\eqref{eq:typeB} and the yield
parameters of~\eqref{eq:typeA}.
But constraining fraction parameters is not equivalent to constraining
yield parameters. If, for example, we know the rate of
a background process as a fraction of the rate of a control process, we
should constrain this fraction. If, on the other hand, we know the absolute
rate, we should constrain the yield.
As pointed out above, the full likelihood
is not required to be Poissonian in its yield parameters.
Thus a Gaussian constraint on a Poissonian yield parameter is the correct implementation,
even if the sum of a Gaussian and a Poissonian random variable does not follow
a Poissonian PDF. The constraint term on
a yield parameter can even have a width smaller than $\sqrt{N}$.
This happens, for example, when the constraint is derived from a large 
control yield $Y\pm \sqrt{Y}$ by
scaling down by a factor $\varepsilon$ that has no uncertainty:
$y_e = \varepsilon Y \pm \varepsilon \sqrt{Y}$.
In such situations,
the constraint term will push the fit into the genuine type~A EML regime.


\section{Pseudo Experiments}
\label{sec:pseudoexps}

Generating and fitting back a large number of pseudo experiments
is a powerful tool to understand and validate a fit procedure. 
Pseudo experiments are generated by drawing a pseudo dataset from the full
PDF, for example through a hit-and-miss algorithm.

In an EML situation it is important that in the pseudo datasets the component 
event yields all fluctuate like a Poissonian. As a consequence, also the total
yield fluctuates like a Poissonian, and each pseudo dataset 
contains a different number of events. Note that each yield must fluctuate
independently, so that their ratios are not constant across the toy
experiments. It is not enough that the total yield fluctuates like a Poissonian. 

If constraints are present, they have to be considered when generating and
fitting a toy dataset.
In particular, there is a ``right'' and a ``wrong way'' of doing it, as outlined
in Ref.~\cite{CDF}. The ``right way'' is to interpret the constraint 
as stemming from an external measurement: We not only have to repeat 
our own measurement (by drawing events from the full PDF), but also have to 
repeat the external measurement by drawing from the constraint PDF. So each
toy experiment will be performed with a different constraint term, but using
the same shape for the total PDF. The ``wrong way'' is to fluctuate 
the total shape and not the constraint term, so that each experiment uses 
the same constraint term, but draws events from different total PDFs. This
will lead to biased results.

If there are constraints present for yield parameters, their correct treatment
in toy generation is still the above ``right way''. This is even though 
the likelihood function does not only contain a Poisson term 
(the EML term), but also a generally non-Poissonian term (the constraint).
Thus one might conclude, that the total yield should not 
be generated from a Poissonian, while this in fact is the case.

Let us be more specific. Fixing the notation, we will denote for a parameter $A$,
its true value as $A_t$, its value as estimated by the fit as $A_f$, its value
as determined by an external measurement as $A_e$, and a generated value as $A'$.
Suppose the total PDF is that of \eqref{eq:typeA},
and we add a Gaussian constraint on the background yield, corresponding to an
external measurement of $N_{b,e} \pm \sigma_{N_{b,e}}$. Obviously the fit will
be biased if we constrain a parameter to anything else but its true value,
so we'll assume $N_{b,e} = N_{b,t}$ (if $N_{b,e}$ was obtained from a genuine
external measurement, this bias will likely go in the direction of the true value). 
To generate a 
toy experiment, we have to
\begin{enumerate}
	\item draw a value $N_b'$ from a Poissonian $P(N; N_{b,t})$,
	and a value $N_s'$ from a Poissonian $P(N; N_{s,t})$,
	\item generate $N_b'$ background events
	from $G(x;\mu_2)$ and $N_s'$ signal events from $G(x;\mu_1)$,
	\item draw a toy constraint value $N_{b,e}'$ from the constraint 
	PDF $G(N_b; N_{b,e}, \sigma_{N_{b,e}})$.
\end{enumerate}
Then the likelihood to be maximized for this particular experiment
is
\begin{equation}
	\mathcal{L}(N_s, N_b) = 
	\frac{e^{-N} N^{N'}}{N'!}
	\cdot G(N_b; N_{b,e}', \sigma_{N_{b,e}})
	\cdot \prod_{i=1}^{N'} g(x'_i)\ ,
\end{equation}
where $N' = N_s' + N_b'$ and $N = N_s + N_b$ is the expected total yield to be estimated
by the fit.

It is interesting to note, that the Poisson EML term is technically
also a constraint. It constrains the fitted total number of events 
to the observed number of events. It also varies with the toy
experiments, because the generated ``observed'' number of events
varies.


\section{Pull Definitions}
\label{sec:pulldefinitions}

The pull statistic is defined as
\begin{equation}
	\label{eq:pull}
	p = \frac{\lambda_f-\lambda_t}{\sigma_{\lambda_f}}\ ,
\end{equation}
where $\lambda_f \pm \sigma_{\lambda_f}$ is the fit result
of one particular pseudo experiment, and $\lambda_t$ is the 
true value. One expects the pull to follow a unit Gaussian, so from its 
observed distribution one can draw conclusions about whether or not the 
fit reports unbiased central values and errors of correct coverage.
If the pull distribution has mean $\mu_p \pm \sigma_{\mu_p}$ and 
width $w_p \pm \sigma_{w_p}$ that
are not equal to 0 and 1, respectively, one can decide to correct the fit result
for these biases:
\begin{align}
	\lambda_f^c &= \lambda_f - \mu_p \sigma_{\lambda_f}\ , \\
	\sigma_{\lambda_f}^c &= w_p \sigma_{\lambda_f}\ .
\end{align}
The pull formed with the corrected quantities then has mean 0 and width 1.


If a constraint to $\lambda_e\pm \sigma_{\lambda_e}$ is present, the usual
pull of \eqref{eq:pull} still follows a unit Gaussian, provided the toy
experiments are generated in the ``right way'' as described in Section~\ref{sec:pseudoexps}.

Reference~\cite{CDF} defines a second pull statistic as
\begin{equation}
	\label{eq:pull2}
	p_2 = \frac{\lambda_f-\lambda_e}{\sqrt{\sigma_{\lambda_e}^2-\sigma_{\lambda_f}^2}} \ .
\end{equation}
The square root is always defined as a consequence of \eqref{eq:weightedaverr}.
Ref.~\cite{CDF} points out that this definition may
exhibit a slower convergence towards the unit Gaussian 
distribution, i.e. for large number of events in the toy samples (not
large number of toy experiments). However, the authors do not discuss constraints 
in the context of EML fits, and we found $p_2$ to not follow a unit
Gaussian even with sufficiently large samples.

A third possibility is
\begin{equation}
	\label{eq:pull3}
	p_3 = \frac{\lambda_f-\lambda_e'}{\sigma_{\lambda_f}}\ ,
\end{equation}
where the generated constraint value $\lambda_e'$ is used rather than the
fixed $\lambda_e$. This definition is used in certain situations by the \textsc{RooFit}
framework~\cite{roofit}, which we will discuss later. When generating
toy experiments in the right way, we found that also $p_3$ does not follow a unit
Gaussian.

Using pull definitions with different convergence rates comes
with an additional complication: If a bias correction is necessary in a
situation with too few events for the limit to be valid, the correction
will depend on the pull definition.

In conclusion, we recommend to use $p_1$ in combination 
with the right way of generating toy experiments. This combination
gives unit pulls even if constraints on yield parameters are present.


\section{Examples}
\label{sec:examples}

\begin{figure}[htb]
\center
\includegraphics[width=0.45\textwidth]{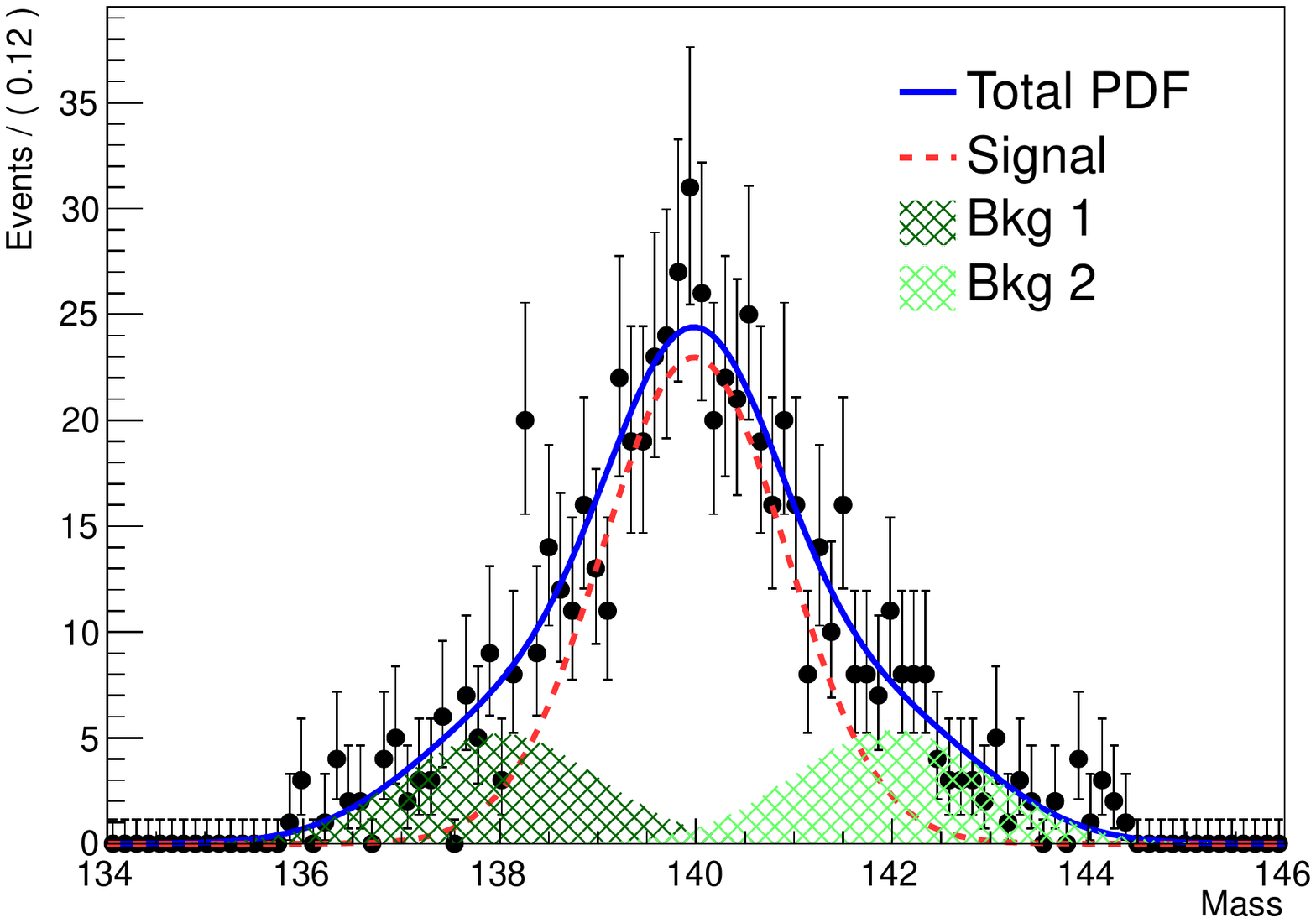}
\caption{Example toy experiment drawn from \eqref{eq:toyexample},
with $N_{s,t}=500$ and $N_{b1,t}=N_{b2,t}=100$.}
\label{fig:toy}
\end{figure}

Let us consider the following example. We will add to the scenario
of \eqref{eq:typeA} a third, low-yield Gaussian, 
to make the situation symmetric. The observable
might represent an invariant mass of a reconstructed composite particle,
and the low-yield Gaussians might correspond to backgrounds, in which
a daughter particle was mis-reconstructed:
\begin{align}
	\label{eq:toyexample}
	g(x) = &N_s G(x;\mu_1=m_0=140) \nonumber \\
	&+ N_{b1} G(x;\mu_{b1}=m_0+2) \nonumber \\
	&+ N_{b2} G(x;\mu_{b2}=m_0-2) \ .
\end{align}
Each Gaussian has unit width.
We will assume the true yields $N_{s,t}=500$ for the signal, and each
$N_{b1,t}=N_{b2,t}=100$ for the backgrounds. We consider Gaussian constraint
terms for both backgrounds $N_{b1}$ and $N_{b2}$,
$G(N_{bi};\mu=N_{bi,t},\sigma=\sqrt{N_{bi,t}})$. An example of such a
pseudo experiment is shown in Fig.~\ref{fig:toy}.

\begin{figure}[htb]
\center
\includegraphics[width=0.45\textwidth]{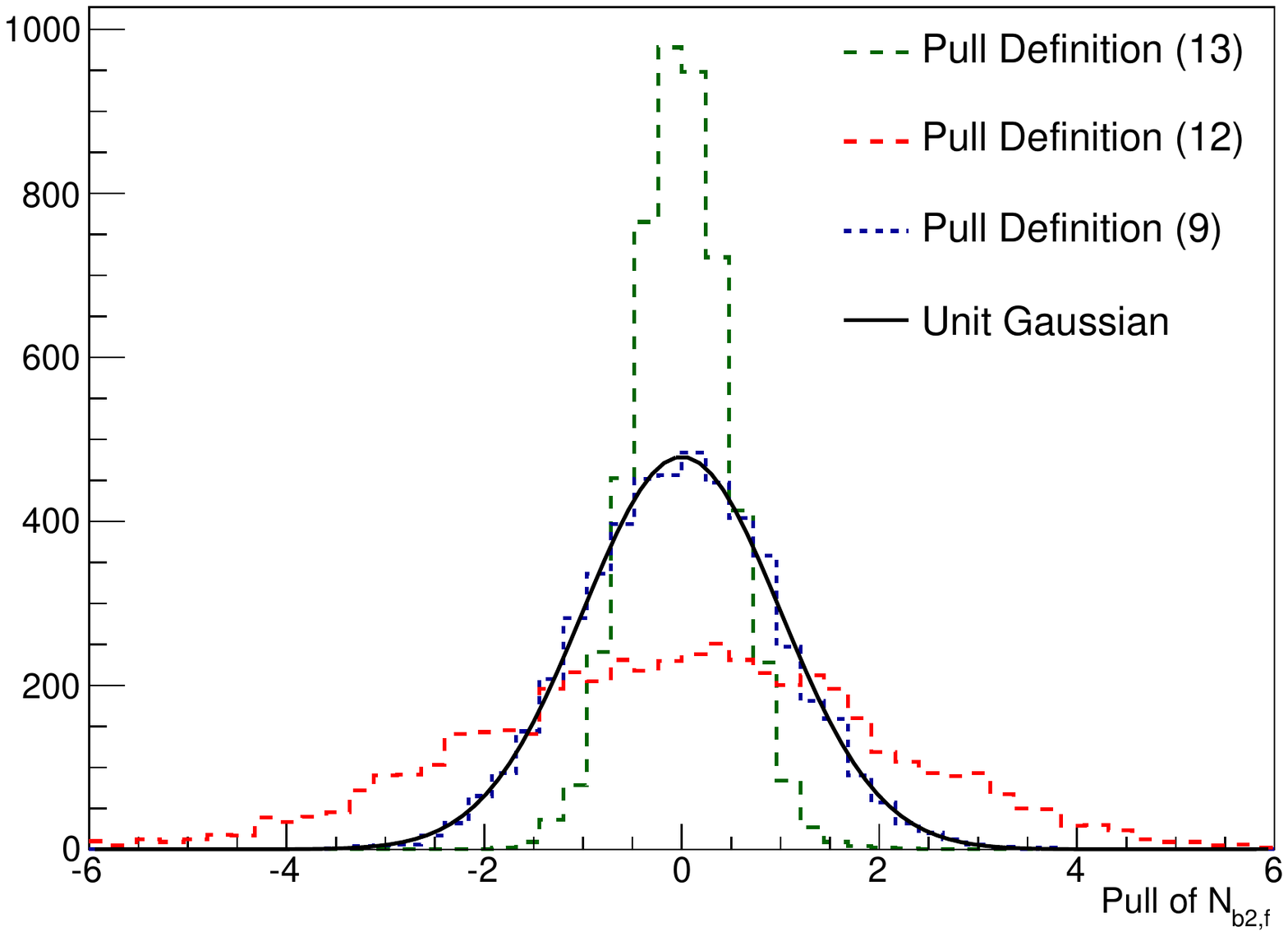}
\caption{Pull distributions according to \eqref{eq:pull} (blue),
\eqref{eq:pull2} (red), and \eqref{eq:pull3} (green). Overlaid
is a unit Gaussian (dashed).}
\label{fig:pullDefs}
\end{figure}

In Figure~\ref{fig:pullDefs} we show the pull distributions of each
pull definition in Section~\ref{sec:pulldefinitions}, using 5000
toy experiments. While the standard definition \eqref{eq:pull} is
consistent with a unit Gaussian ($\mu_p=-0.031\pm0.014$,
$w_p=0.993 \pm 0.011$), the other definitions (\ref{eq:pull2}, \ref{eq:pull3}) are not.
When enlarging the sample sizes to $N_t=70\,000$, the distributions remain
unchanged.

\begin{figure}[htb]
\center
\includegraphics[width=0.45\textwidth]{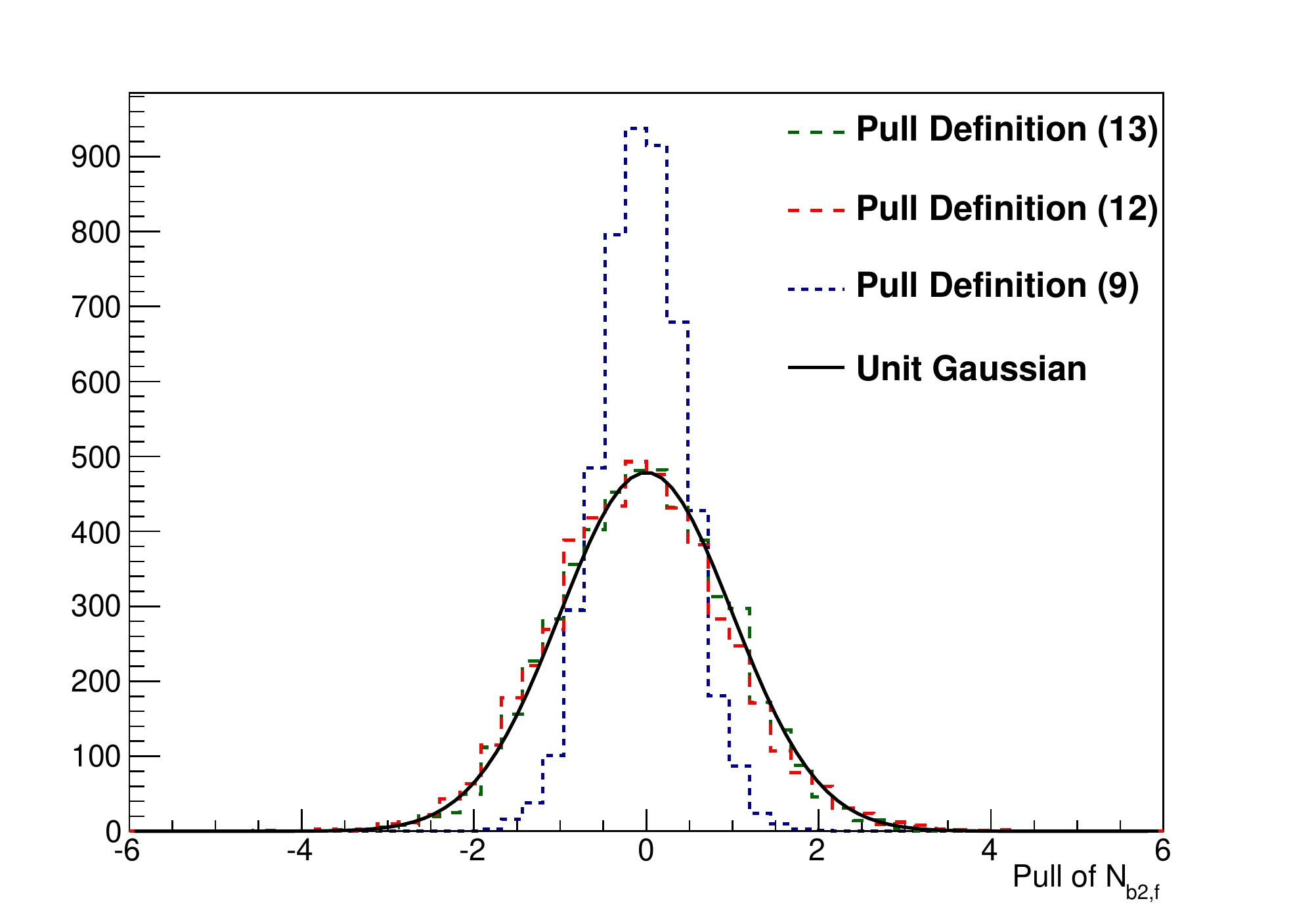}
\caption{Distributions of the same pull definitions as described
in the caption of Figure~\ref{fig:pullDefs}, but when 
performing toy experiments in the ``wrong way''.}
\label{fig:wrongWay}
\end{figure}

In Figure~\ref{fig:wrongWay} we show again the three pull distributions
for generating and fitting the ``wrong way''. Now definitions
\eqref{eq:pull2} and \eqref{eq:pull3} follow a unit Gaussian, while definition 
\eqref{eq:pull} does not. However, this depends on the width of the 
constraint.
These examples support our conclusion of Section~\ref{sec:pulldefinitions}.

%

\begin{figure}[htb]
\center
\includegraphics[width=0.45\textwidth]{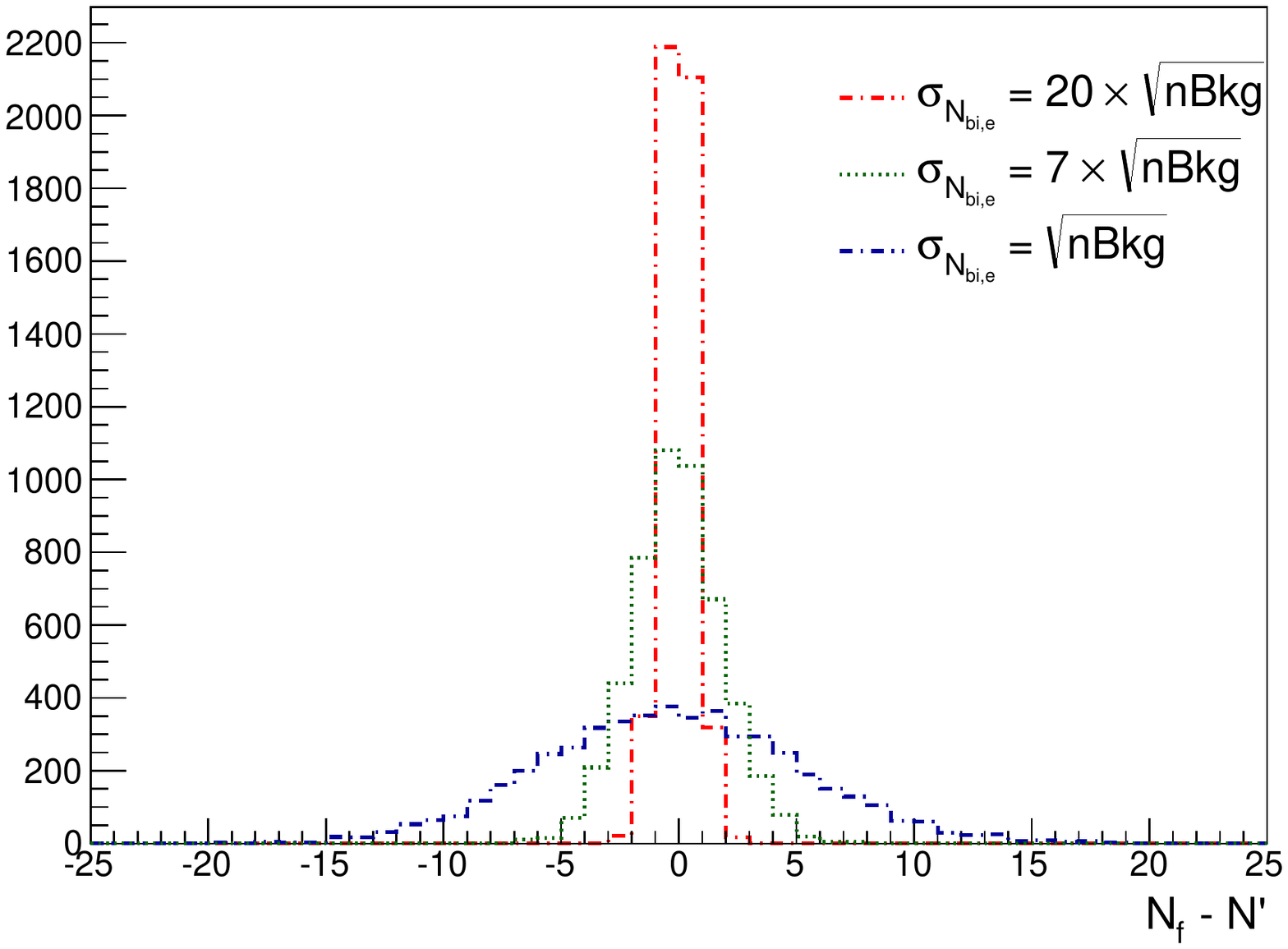}
\caption{Distribution of the difference of fitted
and generated total number of events
$N_f-N'$ for different values of the constraint width
on the $N_{bi}$: $\sigma_{N_{bi,e}}=200$ (red), $70$ (green), $10$
(blue).}
\label{fig:diffNtot}
\end{figure}

To illustrate how a tight constraint term can push the fit into the
type A EML regime, we
now subsequently tighten the constraints on the $N_{bi}$ we observe
in Figure~\ref{fig:diffNtot} that the difference between fitted
and generated total number of events $N_f-N'$ can grow larger,
as the constraints get stronger. The widest distribution
is reached at a value of about $\sqrt{N_{bi,t}}$. Then, deviations of up
to $\approx10$ events are possible, corresponding to $\approx1.4\%$ 
of the events in the considered scenario.

\section{RooFit}

The \textsc{RooFit} framework~\cite{roofit} is widely used in experimental
particle physics to implement sophisticated maximum likelihood fits. It
also features a mechanism to automate pull studies, \texttt{RooMCStudy}.
We would like to point out several pitfalls present
in \textsc{RooFit} version $3.5.4$ (bundled with \textsc{Root} version $5.34.00$).

There are two ways to configure \texttt{RooMCStudy} for the use with
constraint terms. The first, using the \texttt{Constrain()} argument,
is supposed to be used when the constraint term is part of the original PDF
definition. The second, using the \texttt{ExternalConstraints()} argument,
should be used when the constraint terms are supplied separately. Both
ways do not give identical results. In the following, we refer to pulls 
obtained through \texttt{RooMCstudy::plotPull()}.

\begin{figure}[htb]
\center
\includegraphics[width=0.45\textwidth]{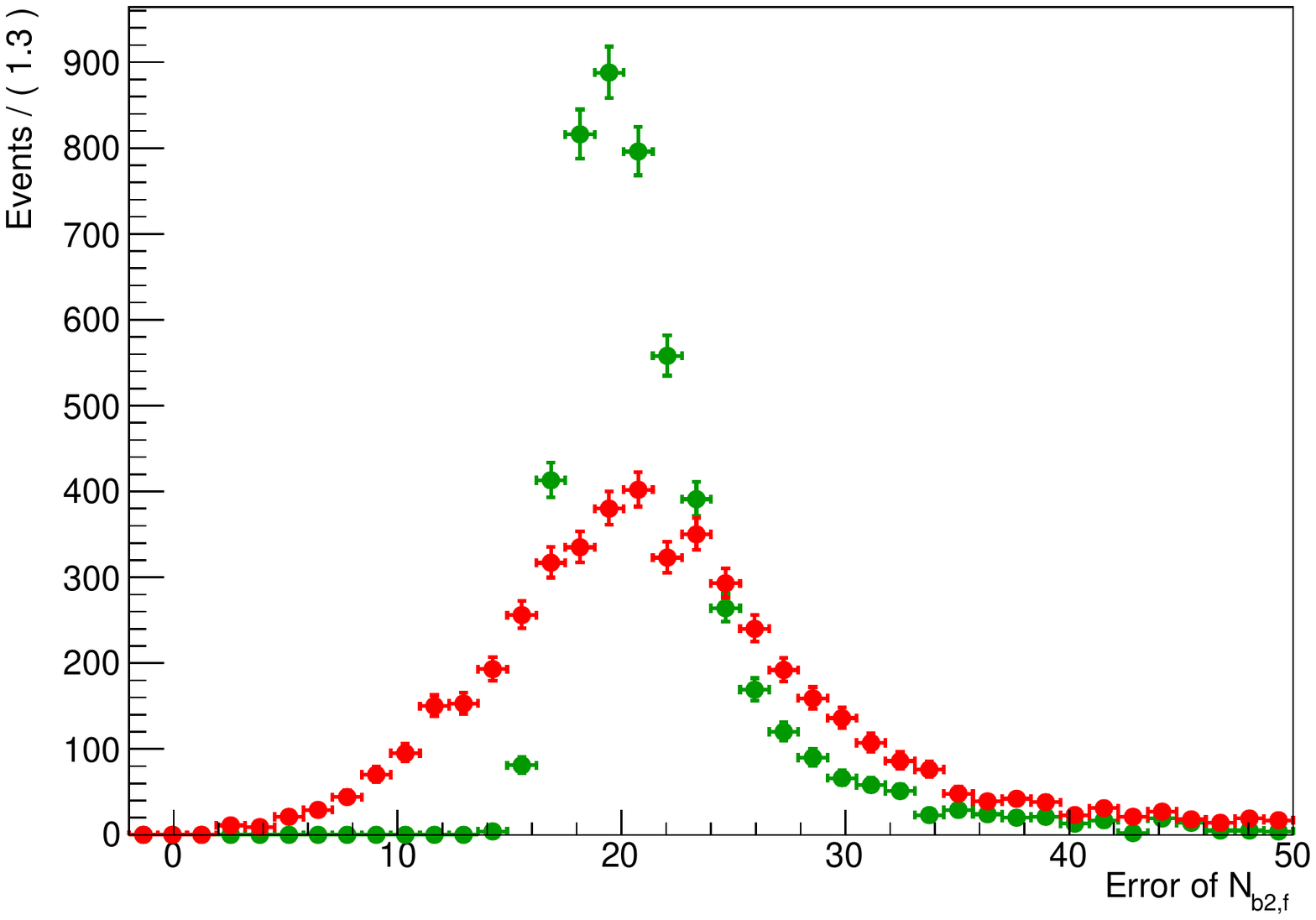}
\caption{Error distribution (red) obtained from \texttt{RooMCstudy} when using
the \texttt{Constrain()} approach and a wide constraint
$G(N_{bi};\mu=N_{bi,t},\sigma=N_{bi,t})$. Also shown is the error
distribution in the unconstrained case (green).}
\label{fig:examples1}
\end{figure}

Using \texttt{Constrain()}: \texttt{RooMCstudy} generates
the ``wrong way'' sketched in Section~\ref{sec:pseudoexps}.  
This is particularly important
if constraints on yield parameters are present. Then, \textsc{RooFit}
first fluctuates the expected yield using the constraint term, and then again
fluctuates the result using the EML Poisson term. As a consequence, the
total generated yield does not follow a Poissonian anymore.
For the pull computation \eqref{eq:pull3} is used.
If the width of a yield constraint is much larger
than $\sqrt{N}$, one expects results similar to those obtained in the unconstrained
situation. But in our example scenario, we observe a moderate bias of $\mu_p=0.3$.
Also, the distributions of both the central value and the error are 
much wider compared to the unconstrained situation. This is shown in 
Figure~\ref{fig:examples1}.
It can also happen, that the effects cancel by chance:
In a second test scenario, corresponding to \eqref{eq:typeA}, with $N_{s,t}=1000$,
$N_{b,t}=500$, and $G(N_b;\mu=N_{b,t}, \sigma=150)$, we observed a unit 
Gaussian pull, while the error and central value distributions of $N_b$
were still too wide.

\begin{figure}[htb]
\center
\includegraphics[width=0.45\textwidth]{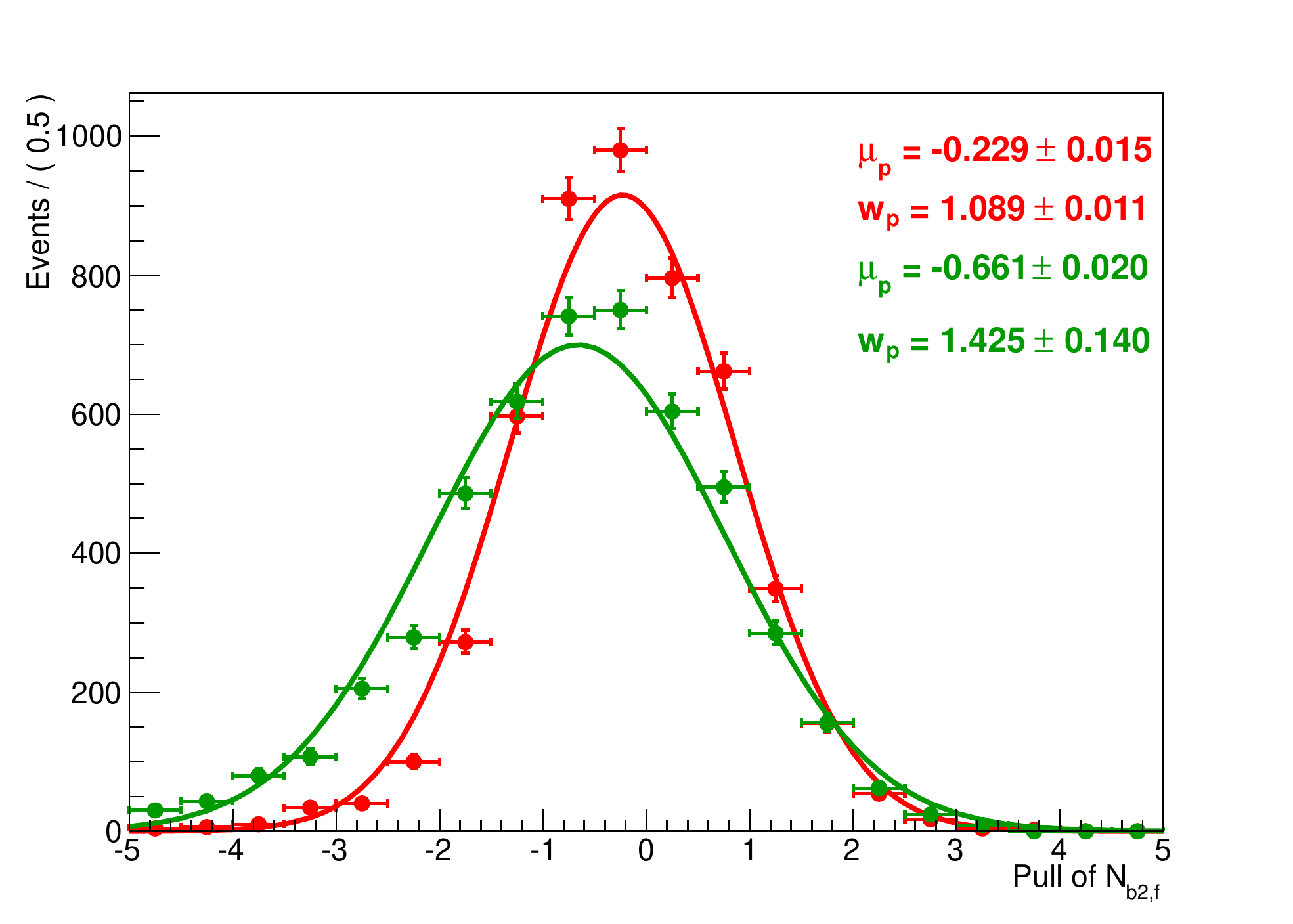}
\caption{Pull distribution (red) obtained from \texttt{RooMCstudy} when using
the \texttt{Constrain()} approach and a wide constraint
$G(N_{bi};\mu=N_{bi,t}=100,\sigma=N_{bi,t})$. The other distribution (green)
is for the same scenario, but obtained by explicitly stating $N_t=700$ in the 
\texttt{generateAndFit()} function call.}
\label{fig:examples2}
\end{figure}

Another pitfall when using \texttt{Constrain()} is that if the 
\texttt{generateAndFit()} function is called in an EML scenario,
and if one explicitly specifies the total number of events to be generated
in the function call, then the pulls depend on the width of the constraint: 
For wide constraints, the pull distribution will be too wide.
This is illustrated in Figure~\ref{fig:examples2}.

\begin{figure}[htb]
\center
\includegraphics[width=0.45\textwidth]{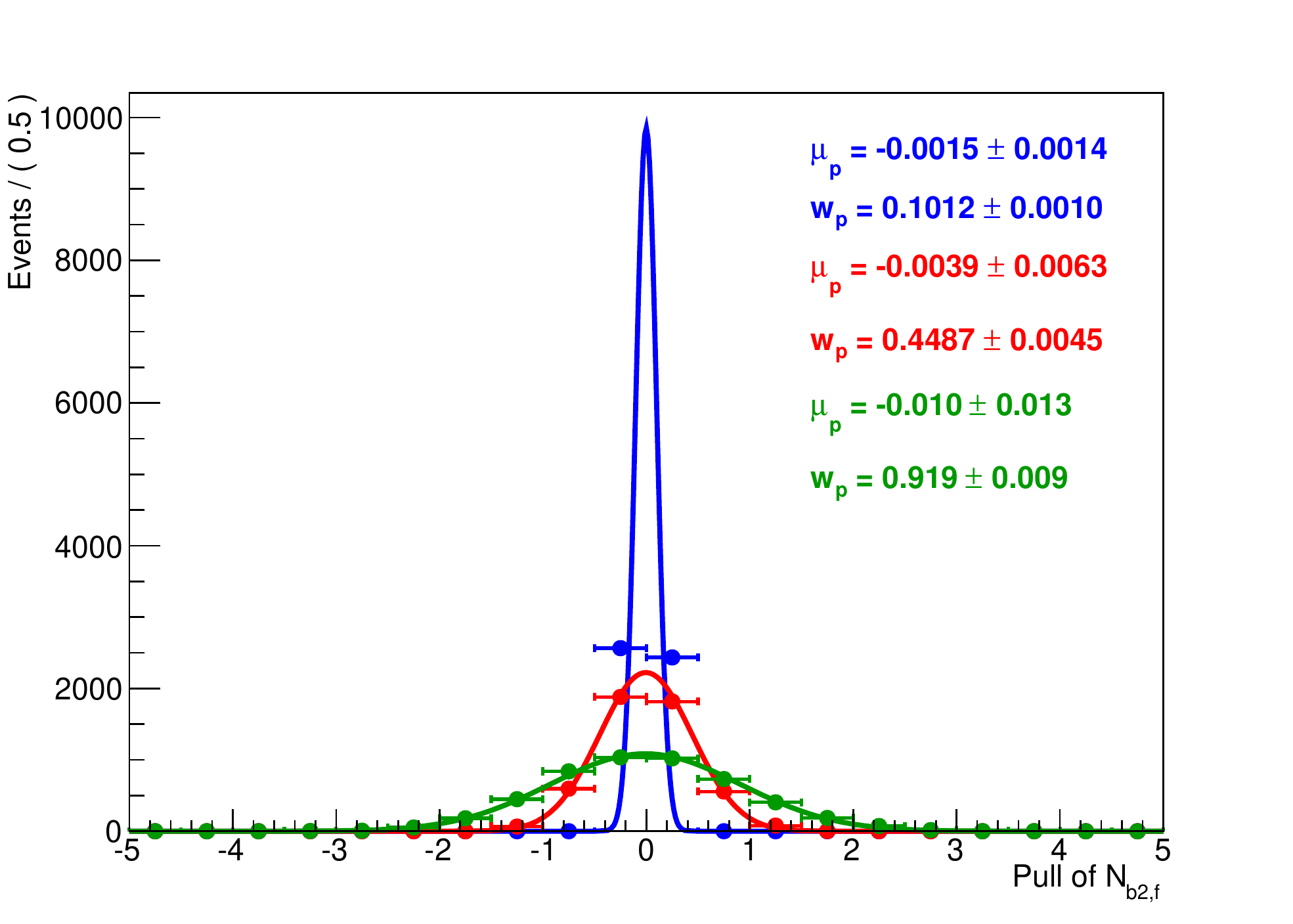}
\caption{Pull distribution obtained from \texttt{RooMCstudy} when using
the \texttt{ExternalConstraints()} approach and a wide constraint
$G(N_{bi};\mu=N_{bi,t}=100,\sigma=50)$ (green), a medium constraint ($\sigma=10$, red),
and a narrow constraint ($\sigma=1$, blue).}
\label{fig:examples3}
\end{figure}

Using \texttt{ExternalConstraints()}: \texttt{RooMCstudy} generates
the ``right way'', i.e. the component yields fluctuate like a Poissonian.
But during fitting, always the same, fixed 
constraint is used, and the pull is computed using \eqref{eq:pull}. As a
consequence, the resulting pull distribution is too narrow. Thus, if the constraint
is wide enough compared to $\sqrt{N}$, the unconstrained situation is recovered.
This is illustrated in Figure~\ref{fig:examples3}.

Considering these difficulties it is clear that, in order to be able
to conclude on a potential fit bias, the user needs a detailed understanding
of \texttt{RooMCstudy}.


\section{Conclusion}

We have discussed the basic features of extended maximum likelihood fits,
and how to use constraint terms to incorporate external knowledge into
these fits. If constraint terms are present, the generation of pseudo
datasets requires care. We recommend to use the ``right way'', in which the constraint
is fluctuated in the generation step and the PDF is not, and to use
the usual pull definition. Then we find the pull to follow a unit
Gaussian even if constraints on yield parameters are present.


The authors wish to thank Niels Tuning
for useful discussion.


\bibliography{Bibliography}{}
\bibliographystyle{h-physrev5}

\end{document}